\documentclass[twocolumn,showpacs,preprintnumbers,amsmath,amssymb]{revtex4}

\usepackage{graphicx}
\usepackage{dcolumn}
\usepackage{bm}
\usepackage{latexsym}
\usepackage{amsfonts}
\usepackage{amssymb}
\usepackage{amsmath}

\begin{document}

\preprint{KUNS-1936}

\title{ On the Validity of a Factorizable Metric Ansatz in String Cosmology }

\author{Sugumi Kanno}
\email{sugumi@tap.scphys.kyoto-u.ac.jp}
\author{Jiro Soda}
\email{jiro@tap.scphys.kyoto-u.ac.jp}
\affiliation{
 Department of Physics,  Kyoto University, Kyoto 606-8501, Japan
}%

\date{\today}

\begin{abstract}
To support the validity of a factorizable metric ansatz used 
 in string cosmology, we investigate a toy problem in RSI model.
 For this purpose, we revise the gradient expansion method
 to conform to the factorizable metric ansatz. 
 By solving the 5-dimensional equations of motion and substituting
 the results into the action, we obtain the 4-dimensional effective 
 action.   It turns out that the resultant action is equivalent to 
 that obtained by assuming the factorizable metric ansatz. 
 Our analysis gives the support of the validity of the factorizable 
 metric ansatz.
\end{abstract}

\maketitle

\section{Introduction}
Results from WMAP strongly support the idea of the inflationary
 universe.  Hence, it is an urgent matter to construct 
  an inflaton potential that agrees with observations based on the
  fundamental theory such as the string theory. 
Recent intense researches on inflationary models 
in string theory have  stemmed from the success to construct
brane inflation with moduli stabilization~\cite{Kachru:2003aw}.
However, almost all studies computing potentials for moduli 
in type IIB string theory suppose a factorizable ansatz for the 
ten dimensional metric:
\begin{eqnarray}
ds^2 = e^{2\omega (y) } g_{\mu\nu} (x) dx^\mu dx^\nu
        + e^{-2\omega (y) + 2u(x) }{\tilde g}_{ab} (y) dy^a dy^b
\end{eqnarray}
where ${\tilde g}_{ab}(y)$ is the metric on the internal Calabi-Yau 
manifold, $\omega (y)$ is the warp factor, 
and $u(x)$ represents the volume modulus. 
This ansatz is based on the static 
 solution~\cite{Giddings:2001yu}
\begin{eqnarray}
ds^2=e^{2\omega (y)}\eta_{\mu\nu}dx^\mu dx^\nu
	+e^{-2\omega (y)}{\tilde g}_{ab}(y)dy^ady^b \ .
\end{eqnarray}
It should be stressed that, in the presence of the moving branes, 
 no proof for the factorizable ansatz exists even at low energy. 
 If this ansatz is not correct, any conclusion derived using it is
 not reliable. In fact, 
 recently, this ansatz is challenged by de Alwis~\cite{deAlwis:2004qh}. 
 As this factorizable ansatz for the metric is crucial in the discussion
of the D-brane inflation, it is important to examine its validity. 

In order to investigate the validity of 
this ansatz, as a modest step,
we focus on the Randall-Sundrum (RS) I model~\cite{RS1}.
 Here, we also have a similar problem.
In considering the cosmology,  
if we follow the factorizable metric ansatz, what we should do is
to replace the Minkowski metric in a static solution
\begin{eqnarray}
ds^2 = a^2 (y) \eta_{\mu\nu} dx^\mu dx^\nu +dy^2 
 \ , \quad a(y)=e^{-y/\ell} 
\end{eqnarray}
with a spacetime dependent metric $g_{\mu\nu}(x)$ as
\begin{eqnarray}
  ds^2 = a^2 (y) g_{\mu\nu} (x) dx^\mu dx^\nu 
                                   + {\cal G}_{yy}(x) dy^2 \ ,
\end{eqnarray}
where we also included the modulus field ${\cal G}_{yy}$.
The question is the validity of this assumption in the context of
the brane cosmology. In this paper, we derive the four dimensional low 
energy effective action  on the brane without using this ansatz. 
 Then, we compare the result with the effective action derived using the
 factorizable ansatz to examine its validity.  
 
 Organization of this paper is as follows. 
 In Sec.II, we present our strategy to attack the issue. 
 In Sec.III, we solve the bulk equations of motion using the revised gradient 
 expansion method. In the Sec.IV, we derive the 4-dimensional effective action
 and  discuss the validity of the factorizable metric ansatz.
 The final section is devoted to the conclusion. 

\section{How to justify the metric ansatz?}
We consider an $S_1/Z_2$ orbifold spacetime with the two branes as the fixed points. In the RSI model, the two flat 3-branes 
are embedded in the 5-dimensional asymptotically anti-deSitter (AdS) bulk 
with the curvature radius $\ell$ with brane tensions given by
$\sigma_+=6/(\kappa^2\ell)$ and $\sigma_-=-6/(\kappa^2\ell)$.
The model is described by the action 
\begin{eqnarray}
S&=&{1\over 2\kappa^2}\int d^5x\sqrt{-\cal G}
	\left[{\cal R}+{12\over\ell^2}\right]\nonumber\\
&&	-{6\over\kappa^2\ell}\int d^4x\sqrt{-g_+}  
	+{6\over\kappa^2\ell}\int d^4x\sqrt{-g_-}\nonumber\\
&&	+{2\over\kappa^2}\int d^4x\sqrt{-g_+}{\cal K}_+
  	-{2\over\kappa^2}\int d^4x\sqrt{-g_-}{\cal K}_-\ , 
      	\label{action:5-dim}
\end{eqnarray}
where $\kappa^2$ is the five-dimensional gravitational coupling 
constant and ${\cal R}$ is the curvature scalar. 
We denoted the induced metric on the positive and negative tension branes 
by $g_{\mu\nu}^+$ and $g_{\mu\nu}^-$, respectively. In the last line, we have 
taken into account the Gibbons-Hawking boundary terms instead of 
introducing delta-function singularities in the curvature.
The factor 2 in the Gibbons-Hawking term comes from the $Z_2$ symmetry of 
this spacetime. ${\cal K}_\pm$ is the trace part of the extrinsic curvature
of the boundary near each one of the branes.

Here, the question is how to obtain the effective action for 
discussing the cosmology. One often takes the metric ansatz
 and substitutes the ansatz into the action to get the 4-dimensional
 effective action.  
Let us assume that the metric is factorizable
\begin{eqnarray}
ds^2= a^2 (y) g_{\mu\nu}(x) dx^\mu dx^\nu + {\cal G}_{yy}(x) dy^2
\label{factor1}
\end{eqnarray}
and the branes are located at the fixed coordinate points. 
 Substituting this metric into the action and integrating out the 
result with respect to $y$,  we obtain the 4-dimensional action.
 However, an inadequate restriction of the functional space in the variational
 problem yields the wrong result. 
The correct procedure to obtain the 4-dimensional effective action is 
  first to solve the bulk equations of motion and substitute the results
 into the original action.  
 To solve the bulk equations, we can employ the gradient expansion method
 ~\cite{KS2,KS,wiseman,sugumi}.
 Our analysis using the gradient expansion method
 shows that the correct metric takes the form~\cite{KS2}:
\begin{eqnarray}
ds^2= a^2 (y \sqrt{{\cal G}_{yy}(x)}) g_{\mu\nu}(x) dx^\mu dx^\nu 
           + {\cal G}_{yy}(x) dy^2
\end{eqnarray}
 which clearly reject the factorizable ansatz (\ref{factor1}). 
 
 However, there is another possibility. 
 One can assume the following factorizable metric
\begin{eqnarray}
  ds^2 = a^2 (y) g_{\mu\nu} (x) dx^\mu dx^\nu + dy^2 \ ,
\end{eqnarray}
and  the branes are moving in the above coordinates.
Namely, the positive and negative tension branes are respectively placed at 
\begin{eqnarray}
y=\phi_+(x),\qquad y= \phi_-(x)  \ ,
\end{eqnarray}
 which are often referred to as the moduli fields. 
 This is  another description of the RSI cosmology and 
 often called as moduli approximation 
in the literature~\cite{Khoury:2001wf}. 
 Although  two scalar fields are introduced, one of which is the extra
degree of freedom as we will see later in Eq.~(\ref{ST}). 
The physical quantity is the difference of these moduli 
fields which corresponds to the radion in our previous work~\cite{KS2}. 
 This ansatz leads to the action~\cite{Khoury:2001wf}
\begin{eqnarray}
S&=&	\frac{\ell}{2\kappa^2}\int d^4x\sqrt{-g}
	\left[\left\{
	a^2(\phi_+)-a^2(\phi_-)\right\}R(g) \right.\nonumber\\
&&\quad
	\left.
	+\frac{6}{\ell^2}\left\{
	a^2(\phi_+)(\partial\phi_+)^2
	-a^2(\phi_-)(\partial\phi_-)^2
	\right\}\right] \ .
	\label{moduliapp}
\end{eqnarray}
In this  case, we do not have the result to be compared,
because the gradient expansion method is not prepared for this
 parameterization of the model. 
Our aim is to examine the validity of this ansatz
 by conforming the gradient expansion method to the 
 factorizable metric ansatz.

\section{Revised Gradient Expansion Method}

The metric we take in solving the bulk equations of motion
is the one in the Gaussian normal coordinate system
\begin{eqnarray}
ds^2=\gamma_{\mu\nu}(y,x)dx^\mu dx^\nu+dy^2 \ ,
\end{eqnarray}
where the factorized metric is not assumed. 

Now we give the basic equations in the bulk.
When solving the bulk equations of motion, it is convenient to define 
the extrinsic curvature on the $y = {\rm constant} $ slicing as
$K_{\mu\nu} 
     = - {1\over 2} {\partial \over \partial y} 
                              \gamma_{\mu\nu}  $.
Decomposing this extrinsic curvature into the traceless part and the trace part
\begin{equation}
K_{\mu\nu}=\Sigma_{\mu\nu}+{1\over 4}\gamma_{\mu\nu}K  \ , \quad
K = - {\partial \over \partial y}\log \sqrt{-\gamma}    \  ,
\label{K} 
\end{equation}
we obtain the basic equations which hold in the bulk;
\begin{eqnarray}
&&\Sigma^\mu{}_{\nu ,y}-K\Sigma^\mu{}_{\nu} 
	=-\left[
	R^\mu{}_\nu(\gamma) 
	-{1\over 4} \delta^\mu_\nu R(\gamma) 
        \right]
        \label{munu-trc} \ ,     \\
&&{3\over 4}K^2-\Sigma^\alpha{}_{\beta}\Sigma^\beta{}_{\alpha} 
	=R(\gamma)
	+{12\over\ell^2}
	\label{munu-trclss} \ ,  \\
&&\nabla_\lambda\Sigma_{\mu}{}^{\lambda}  
	-{3\over 4}\nabla_\mu K = 0
	\label{ymu} \ ,
\end{eqnarray}
where $\nabla_\mu $ denotes the covariant derivative with respect to 
the metric $\gamma_{\mu\nu}$ and $R^\mu{}_\nu(\gamma)$ is the corresponding 
curvature.

The effective action have to be derived by substituting the solution of
 Eqs.~(\ref{munu-trc})$\sim$(\ref{ymu}) 
into the action (\ref{action:5-dim}) and 
integrating out the result over the bulk coordinate $y$. 
In reality, it is difficult to perform this general procedure.
However, what we need is  the low energy effective theory. 
At low energy, the energy density of the matter, $\rho$, 
on a brane is smaller than the brane tension, i.e.,
$\rho /|\sigma| \ll 1$. In this regime, the 4-dimensional curvature
can be neglected compared with the extrinsic curvature. 
Thus, the Anti-Newtonian or gradient expansion method used in the cosmological 
context~\cite{tomita} is applicable to our problem. 

\subsection{Zeroth Order}

At zeroth order,  we can neglect the curvature term in 
Eqs.~(\ref{munu-trc})$\sim$(\ref{ymu}). 
Moreover, the tension term only induces the isotropic bending of the brane.
Thus, an anisotropic term vanishes at this order,  
$\overset{(0)}{\Sigma}{}^\mu{}_\nu=0$. 
As the result,  we obtain
\begin{equation}
     \overset{(0)}{K} = {4\over \ell}  \quad {\rm or} \quad
     \overset{(0)}{K}{}^\mu{}_{\nu} = {1\over\ell}
	\delta^{\mu}_{\nu} \ .
\end{equation}
Using the definition of the extrinsic curvature
\begin{equation}
     \overset{(0)}{K}{}_{\mu\nu} 
     = - {1\over 2} {\partial \over \partial y} 
                               \overset{(0)}{\gamma}{}_{\mu\nu}     \   ,
\end{equation}
we get the zeroth order metric  as
\begin{equation}
ds^2 = dy^2 +  a^2 (y) g_{\mu\nu}(x) dx^\mu dx^\nu\ ,
\quad
a(y)=e^{-y/\ell}    \ ,
\end{equation}
where  the tensor $g_{\mu\nu}$ is  the constant of integration
 which weakly depends on the brane coordinates $x^\mu$.

\subsection{First Order}

Our iteration scheme is to write the metric $\gamma_{\mu\nu}$ 
as a sum of local tensors built out of $g_{\mu\nu}$, 
with the number of derivatives increasing with the order
of iteration, that is, 
$ O((\ell/L)^{2n})$, $n=0,1,2,\cdots$. Here, $L$ represents
the characteristic length scale of the 4-dimensional curvature.
Hence, we seek the metric as a perturbative series 
\begin{eqnarray}
&&\gamma_{\mu\nu} (y,x) =
	a^2(y)\left[ g_{\mu\nu} (x) 
  	+f_{\mu\nu}(y,x)
      	+ \cdots  \right]  \ .
      	\label{expansion:metric}
\end{eqnarray}
The effective action can be constructed with the knowledge of
 the  leading order metric $f_{\mu\nu}(y,x)$. 
 Other quantities can be also expanded as
\begin{eqnarray}
K^\mu{}_{\nu}&=&
	{1\over\ell}
	\delta^{\mu}_{\nu}
        +\overset{(1)}{K}{}^{\mu}{}_{\nu}
	+\overset{(2)}{K}{}^{\mu}{}_{\nu}+\cdots  \ , \nonumber\\
\Sigma^\mu{}_{\nu}
	&=&  \ \  0 \ \ 
	+\overset{(1)}{\Sigma}{}^{\mu}{}_{\nu}
	+\overset{(2)}{\Sigma}{}^{\mu}{}_{\nu} + \cdots          \ .
\end{eqnarray}

The first order solutions are obtained by taking into account the 
terms neglected at zeroth order. 
At  first order,  Eqs.~(\ref{munu-trc})$\sim$(\ref{ymu}) become
\begin{eqnarray}
&&\overset{(1)}{\Sigma}{}^{\mu}{}_{\nu , y} 
	-{4\over\ell} \overset{(1)}{\Sigma}{}^{\mu}{}_{\nu} 
	=-\left[R^\mu{}_\nu(\gamma) 
	-{1\over 4} \delta^\mu_\nu R(\gamma)\right]^{(1)}
	\label{1:munu} \ , \\
&&{6 \over\ell} \overset{(1)}{K} = \left[~R(\gamma)
	~\right]^{(1)}
	\label{1:trace} \  ,\\
&&\overset{(1)}{\Sigma}{}_{\mu}{}^{\lambda}{}_{|\lambda}  
	-{3\over 4}\overset{(1)}{K}{}_{|\mu} = 0 
	\label{1:ymu}\ .
\end{eqnarray}
where the superscript $(1)$ represents the order of the derivative expansion
 and $|$ denotes the covariant derivative with respect to
 the metric $g_{\mu\nu}$.
Here, $[R^\mu{}_\nu(\gamma)]^{(1)} $ 
means that the curvature is approximated by
 taking the Ricci tensor of $a^2(y)g_{\mu\nu}(x)$ 
 in place of $R^{\mu}{}_{\nu}(\gamma)$. 
 It is also convenient to write it in terms of the Ricci
  tensor of $g_{\mu\nu}$, denoted by $R^\mu{}_\nu (g)$.
 
Substituting the zeroth order metric into $R(\gamma)$, 
we can write Eq.~(\ref{1:trace}) as
\begin{equation}
\overset{(1)}{K} = {\ell\over 6a^2} R(g)
\label{1:trc} \ .
\end{equation}
Hereafter, we omit the argument of the curvature for simplicity. 
Simple integration of Eq.~(\ref{1:munu}) also gives the traceless part
 of the extrinsic curvature as
\begin{equation}
\overset{(1)}{\Sigma}{}^{\mu}{}_{\nu}={\ell\over 2a^2}
	(R^\mu{}_{\nu}-{1\over 4}\delta^\mu_\nu R)  
	+{\chi^{\mu}{}_{\nu}(x)\over a^4}
	\label{1:trclss}  \ ,
\end{equation}
where $\chi^\mu{}_\nu$ is the constant of integration which satisfies 
\begin{eqnarray}
\chi^{\mu}{}_{\mu}=0   \ , \quad\chi^{\mu}{}_{\nu|\mu}=0 \ .
\label{TT}
\end{eqnarray}
Here, the latter condition came from Eq.~(\ref{1:ymu}). 
In our previous work~\cite{KS2}~\cite{KS}, we find this term corresponds 
to the dark  radiation at this order.  
 Due to the traceless property, $\chi^\mu{}_{\nu}$ is not relevant to the 
 derivation of the effective action. 

 From Eqs.(\ref{1:trc}) and  (\ref{1:trclss}),
the correction to the metric $g_{\mu\nu}$ at this order can be obtained as
\begin{eqnarray}
f_{\mu\nu}(y,x^\mu)&=&-{\ell^2\over 2a^2} 
	\left( R_{\mu\nu}-{1\over 6} g_{\mu\nu} R \right)
	\nonumber\\
&&\qquad\qquad	
-{\ell \over 2a^4}\chi_{\mu\nu}+C_{\mu\nu}(x) \ ,
\end{eqnarray}
where $C_{\mu\nu}$ is the constant of integration which 
will be  fixed later in Eq.~(\ref{C}). 

\section{Effective Action}

Now, up to the first order, we have
\begin{eqnarray}
g_{\mu\nu}(y,x)=a^2(y)
	\left[
	g_{\mu\nu}(x)+f_{\mu\nu}(y,x)
	\right]    \ . 
\end{eqnarray}
In the following, we will calculate the bulk action, $S_{\rm bulk}$, 
the actions for each brane, $S_\pm$ and the 
Gibbons-Hawking term, $S_{\rm GH}$, separately. After that, we collect all of 
them and obtain the 4-dimensional effective action.

In order to calculate the bulk action, 
we need the determinant of the bulk metric 
\begin{eqnarray}
\sqrt{\cal -G}&=&a^4(y)\sqrt{-g}\sqrt{1-\frac{\ell^2}{6a^2}R+C^\mu{}_\mu}
	\nonumber\\
&\approx&
	a^4(y)\sqrt{-g}\left(
	1-\frac{\ell^2}{12a^2}R\right)
	\left(1+\frac{C^\mu{}_\mu}{2}\right)   \ , 
\end{eqnarray}
where we neglected the second order quantities. 
Then the bulk action becomes
\begin{eqnarray}
S_{\rm bulk}&\equiv&{1\over 2\kappa^2}\int d^5x\sqrt{\cal -G}
	\left[{\cal R}+{12\over\ell^2}\right]
	\nonumber\\
&=&-\frac{8}{\kappa^2\ell^2}\int d^4x\sqrt{-g}
	\left[\frac{\ell}{4}\left\{
	a^4(\phi_+)-a^4(\phi_-)\right\}
	\right.\nonumber\\
&&	\left.
	-\frac{\ell^3}{24}\left\{
	a^2(\phi_+)-a^2(\phi_-)\right\}R~
	\right]\left[1+\frac{C^\mu{}_\mu}{2}\right] \ ,
	\label{bulk}
\end{eqnarray}
where we have used  the 
the equation ${\cal R} = -20/\ell^2$ which holds in the bulk. 
Notice that the Ricci scalar came from ${\rm tr}f_{\mu\nu}$ 
in $\sqrt{\cal -G}$.

Next, let us calculate the action for the brane tension. 
 The induced metric on each 
brane is  written by
\begin{eqnarray}
g^\pm_{\mu\nu}(\phi_\pm,x)=
	a^2g_{\mu\nu}(x)+a^2f_{\mu\nu}(\phi_\pm,x)
	+\partial_\mu\phi_\pm\partial_\nu\phi_\pm  \ .
\end{eqnarray}
The determinant of the induced metric can be calculated as 
\begin{eqnarray}
\sqrt{-g_\pm}&=&a^4(\phi_+)\sqrt{-g}
	\sqrt{1+\frac{1}{a^2}(\partial\phi_\pm)^2
	-\frac{\ell}{6a^2}R+C^\mu{}_\mu}
	\nonumber\\
&\approx&
	a^4(\phi_+)\sqrt{-g}\left(
	1+\frac{1}{2a^2}(\partial\phi_\pm)^2
	-\frac{\ell}{12a^2}R\right)
	\nonumber\\
	&&\times\left(1+\frac{C^\mu{}_\mu}{2}\right) \ ,
\end{eqnarray}
where $(\partial\phi_\pm)^2$ means 
$\partial^\alpha\phi_\pm\partial_\alpha\phi_\pm$. 
Thus, the action for each brane becomes
\begin{eqnarray}
S_\pm &\equiv& \mp {6\over\kappa^2\ell}\int d^4x\sqrt{-g_\pm}  
	\nonumber\\
&=& \mp {6\over\kappa^2\ell}\int d^4x\sqrt{-g}
	\left[a^4(\phi_\pm)
	+\frac{a^2(\phi_\pm)}{2}(\partial\phi_\pm )^2
	\right.\nonumber\\
&&	\left.
	-\frac{\ell^2}{12}a^2(\phi_\pm)~R~\right]\left[
	1+\frac{C^\mu{}_\mu}{2}\right]  \ .
	\label{tension}
\end{eqnarray}
Note that the Ricci scalar came from ${\rm tr}f_{\mu\nu}$
in $\sqrt{-g_\pm}$.

In order to calculate the Gibbons-Hawking term, we need the extrinsic
curvature defined by
\begin{eqnarray}
{\cal K}_{\mu\nu}\equiv n_A\left(
	\frac{\partial^2x^A}{\partial\xi^\mu\partial\xi^\nu}
	\right)
	+\Gamma^A_{BD}
	\frac{\partial x^B}{\partial\xi^\mu}
	\frac{\partial x^D}{\partial\xi^\nu}  \ ,
\end{eqnarray}  
where $x^A$ is the coordinate of the brane, $\xi^\mu = x^\mu $ 
is the one on the brane 
and $n_A$ is the normal vector to the brane. Note that ${\cal K}_{\mu\nu}$ 
is different from $K_{\mu\nu}$ in Eq.~(\ref{K}). The Christoffel symbols 
we need are
\begin{eqnarray}
\Gamma^y_{\mu\nu}&=&\frac{1}{\ell}a^2\left(
	g_{\mu\nu}+f_{\mu\nu}\right)
	-\frac{1}{2}a^2f_{\mu\nu,y}\ \ , \\
\Gamma^\alpha_{y\mu}&=&-\frac{1}{\ell}\delta^\alpha_\mu
	+\frac{1}{2}g^{\alpha\beta}f_{\beta\mu,y} \ .
\end{eqnarray}
The tangent basis on the brane are given by
\begin{eqnarray}
\frac{\partial x^A}{\partial\xi^\mu}
=(\delta^\alpha_\mu,\partial_\mu\phi_\pm) \ .
\end{eqnarray}
Thus, the normal vector takes the form 
\begin{eqnarray}
n_A=(-n_y\partial_\alpha\phi_\pm,n_y) \ .
\end{eqnarray}
From the normalization condition $n_A n^A =1$, we have
\begin{eqnarray}
n_y=\frac{1}{\sqrt{1
	+\frac{1}{a^2}(\partial\phi_\pm)^2}} \ .
\end{eqnarray}
Then the extrinsic curvature is calculated as
\begin{eqnarray}
{\cal K}^\pm_{\mu\nu}&=&n_y\left[
	\nabla_\mu\nabla_\nu\phi_\pm
	+\frac{a^2}{\ell}\left(
	g_{\mu\nu}+f_{\mu\nu}-\frac{\ell}{2}f_{\mu\nu,y}
	\right)\right.\nonumber \\
&&	\left.\qquad
	+\frac{2}{\ell}\partial_\mu\phi_\pm\partial_\nu\phi_\pm
	\right]     \ .
\end{eqnarray}
The trace part of extrinsic curvature on each brane is
\begin{eqnarray}
{\cal K}_\pm&=&g^{\mu\nu}_\pm{\cal K}^\pm_{\mu\nu}\nonumber\\
	&=&n_y\left[
	\frac{4}{\ell}+\frac{1}{a^2}\square\phi_\pm
	+\frac{1}{\ell a^2}(\partial\phi_\pm)^2
	+\frac{\ell}{6a^2}R
	\right] \ .
\end{eqnarray}
Therefore, the Gibbons-Hawking term is obtained as
\begin{eqnarray}
S_{\rm GH}&\equiv&{2\over\kappa^2}\int d^4x\sqrt{-g_+}{\cal K}_+
	-{2\over\kappa^2}\int d^4x\sqrt{-g_-}{\cal K}_-
	\nonumber\\
&=&	{2\over\kappa^2}\int d^4x\sqrt{-g}\left[
	\frac{4}{\ell}a^4(\phi_+)+\frac{3}{\ell}a^2(\phi_+)
	(\partial\phi_+)^2
	\right.\nonumber\\
&&	\left.
	-\frac{\ell}{6}a^2(\phi_+)~R~\right]\left[
	1+\frac{C^\mu{}_\mu}{2}\right]
	-\left(\phi_+\rightarrow\phi_-\right) \ .
	\label{GH}
\end{eqnarray}
Note that the Ricci scalar came from ${\rm tr}f_{\mu\nu}$ in
$\sqrt{-g_\pm}$ and ${\rm tr}f_{\mu\nu,y}$ in ${\cal K}_\pm$.

Substituting the results Eqs.~(\ref{bulk}), (\ref{tension}) and (\ref{GH})
into the 5-dimensional action Eq.~(\ref{action:5-dim}), 
we get the 4-dimensional effective action
\begin{eqnarray}
S&=&S_{\rm bulk}+S_++S_-+S_{\rm GH}\nonumber\\
&=&	\frac{\ell}{2\kappa^2}\int d^4x\sqrt{-g}
	\left[\left\{
	a^2(\phi_+)-a^2(\phi_-)\right\}R \right.\nonumber\\
&&\quad
	\left.
	+\frac{6}{\ell^2}\left\{
	a^2(\phi_+)(\partial\phi_+)^2
	-a^2(\phi_-)(\partial\phi_-)^2
	\right\}\right]\nonumber\\
&&\quad
	\times\left[
	1+\frac{C^\mu{}_\mu}{2}\right]   \ .
	\label{4d}
\end{eqnarray}
Here, $C^\mu{}_\mu$ is the first order quantity, so we can ignore
this term at leading order. We see that this effective action (\ref{4d})
is indistinguishable from Eq.~(\ref{moduliapp}) obtained by assuming the 
factorizable metric.
Thus, we have shown that the action
obtained from the factorizable ansatz is correct at the leading order.

 Here,
it should be stressed that the Einstein-Hilbert term is originated from
the contributions of $f_{\mu\nu}$ in each $S_{\rm bulk}, 
S_\pm$ and $S_{\rm GH}$, so the correction $f_{\mu\nu}$ to the 
 metric $g_{\mu\nu}$ plays an important role. 

 Note that the induced metric on the positive tension brane is
\begin{eqnarray}
g^+_{\mu\nu}(\phi_+,x)&=&a^2(\phi_+)g_{\mu\nu}(x)\ ,
\end{eqnarray}
where we have chosen the constant of integration $C_{\mu\nu}$ to be
\begin{eqnarray}
f_{\mu\nu}(\phi_+,x)=-\frac{1}{a^2(\phi_+)}\partial_\mu\phi_+\partial_\nu\phi_+
\label{C}
\ .
\end{eqnarray}
We see that the induced metric on the positive tension brane is 
different from the factorized metric $g_{\mu\nu}$. Using a conformal
transformation: $g_{\mu\nu}=(1/a^2(\phi_+))g^+_{\mu\nu}$ to rewrite 
the effective action in terms of the induced metric, we finally get
\begin{eqnarray}
S&=&\frac{\ell}{2\kappa^2}\int d^4x\sqrt{-g_+}
	\left[\left\{1-\left(\frac{a(\phi_-)}{a(\phi_+)}\right)^2\right\}
	R(g_+)\right.\nonumber\\
&&\qquad\quad\left.
	-6~\partial_\mu\left(\frac{a(\phi_-)}{a(\phi_+)}\right)
	\partial^\mu\left(\frac{a(\phi_-)}{a(\phi_+)}\right)
	\right]   \ ,
	\label{ST}
\end{eqnarray}
where
\begin{eqnarray}
\frac{a(\phi_-)}{a(\phi_+)}={\rm exp}\left[-\frac{1}{\ell}
	\left(\phi_--\phi_+\right)\right]\ .
\end{eqnarray}
Two moduli fields appear only in the form of the difference,
which corresponds to the radion field. In physical frame, 
the extra degree of freedom disappears.

\section{Conclusion}
 To support the validity of the factorizable metric ansatz used 
 in string cosmology, we investigated a toy problem in RSI model.
 For this purpose, we have revised the gradient expansion method
 to conform to the the factorizable metric ansatz. 
 We have solved the 5-dimensional equations of motion and substituted
 the results into the action. Consequently, we have obtained the
 4-dimensional effective action which is equivalent to that obtained by
 assuming the factorizable metric ansatz. Hence, our calculation supports
 the factorizable metric ansatz. 
 
In the higher order analysis including Kaluza-Klein corrections, 
 the factorizable metric ansatz cannot be correct anymore~\cite{sugumi}. 
 However, in string cosmology, what 
we want is the leading order action. Then the factorizable metric 
ansatz becomes useful method when discussing the cosmology without 
solving the bulk equations of motion at least at the leading order.

Finally, we should mention the limitation of our approach.
Although we have shown the validity of the factorizable ansatz in 
5-dimensions, the issue is still unclear in the case of higher codimension. 
This is because the higher codimension objects is difficult to treat
in a relativistic manner.  
To give a more strong support of the factorizable metric ansatz,
 we need to settle this issue.

\begin{acknowledgements}
This work was supported in part by  Grant-in-Aid for  Scientific
Research Fund of the Ministry of Education, Science and Culture of Japan 
 No. 155476 (SK) and  No.14540258 (JS) and also
  by a Grant-in-Aid for the 21st Century COE ``Center for
  Diversity and Universality in Physics".  
\end{acknowledgements}

%


\begin{thebibliography}{99}

\bibitem{Kachru:2003aw}
S.~Kachru, R.~Kallosh, A.~Linde and S.~P.~Trivedi,
Phys.\ Rev.\ D {\bf 68}, 046005 (2003)
[arXiv:hep-th/0301240];
S.~Kachru, R.~Kallosh, A.~Linde, J.~Maldacena, L.~McAllister and S.~P.~Trivedi,
JCAP {\bf 0310}, 013 (2003)
[arXiv:hep-th/0308055];
E.~Silverstein and D.~Tong,
arXiv:hep-th/0310221;
M.~Alishahiha, E.~Silverstein and D.~Tong,
arXiv:hep-th/0404084.

\bibitem{Giddings:2001yu}
S.~B.~Giddings, S.~Kachru and J.~Polchinski,
Phys.\ Rev.\ D {\bf 66}, 106006 (2002)
[arXiv:hep-th/0105097].

\bibitem{deAlwis:2004qh}
S.~P.~de Alwis,
arXiv:hep-th/0407126.

\bibitem{RS1}
L.~Randall and R.~Sundrum,
Phys.\ Rev.\ Lett.\  {\bf 83}, 3370 (1999)
[arXiv:hep-ph/9905221];

\bibitem{KS2}
S.~Kanno and J.~Soda,
Phys.\ Rev.\ D {\bf 66}, 083506 (2002)
[arXiv:hep-th/0207029];

\bibitem{KS}
S.~Kanno and J.~Soda,
Phys.\ Rev.\ D {\bf 66}, 043526 (2002)
[arXiv:hep-th/0205188];
S.~Kanno and J.~Soda,
Astrophys.\ Space Sci.\  {\bf 283}, 481 (2003)
[arXiv:gr-qc/0209087];
S.~Kanno and J.~Soda,
Gen.\ Rel.\ Grav.\  {\bf 36}, 689 (2004)
[arXiv:hep-th/0303203].

\bibitem{wiseman}
T.~Chiba,
Phys.\ Rev.\ D {\bf 62}, 021502 (2000)
[arXiv:gr-qc/0001029];
T.~Wiseman,
Class.\ Quant.\ Grav.\  {\bf 19}, 3083 (2002)
[arXiv:hep-th/0201127];
T.~Shiromizu and K.~Koyama,
Phys.\ Rev.\ D {\bf 67}, 084022 (2003)
[arXiv:hep-th/0210066].

\bibitem{sugumi}
S.~Kanno and J.~Soda,
 Phys. Lett. B 588, 203-209 (2004)
[arXiv:hep-th/0312106];
P.~L.~McFadden and N.~Turok,
arXiv:hep-th/0409122;
S.~Kanno and J.~Soda,
arXiv:hep-th/0407184.

\bibitem{Khoury:2001wf}
J.~Khoury, B.~A.~Ovrut, P.~J.~Steinhardt and N.~Turok,
Phys.\ Rev.\ D {\bf 64}, 123522 (2001)
[arXiv:hep-th/0103239].
J.~Garriga, O.~Pujolas and T.~Tanaka,
Nucl.\ Phys.\ B {\bf 655}, 127 (2003)
[arXiv:hep-th/0111277];
P.~Brax, C.~van de Bruck, A.~C.~Davis and C.~S.~Rhodes,
Phys.\ Rev.\ D {\bf 67}, 023512 (2003)
[arXiv:hep-th/0209158];
G.~A.~Palma and A.~C.~Davis,
arXiv:hep-th/0406091;
S.~L.~Webster and A.~C.~Davis,
arXiv:hep-th/0410042.

\bibitem{tomita}
K.~Tomita,
Prog.\ Theor.\ Phys.\  {\bf 54}, 730 (1975);
G.~L.~Comer, N.~Deruelle, D.~Langlois and J.~Parry,
Phys.\ Rev.\ D {\bf 49}, 2759 (1994);
D.~S.~Salopek and J.~M.~Stewart,
Phys.\ Rev.\ D {\bf 47}, 3235 (1993);
J.~Soda, H.~Ishihara and O.~Iguchi,
Prog.\ Theor.\ Phys.\  {\bf 94}, 781 (1995)
[arXiv:gr-qc/9509008].




\end{thebibliography}

\end{document}